\documentclass[12pt]{JHEP3}
\usepackage{setspace}
\usepackage[dviwin]{graphicx}
\usepackage{amsfonts}
\usepackage{amssymb}
\usepackage{mathrsfs}
\usepackage{amsmath}
\usepackage{epsfig}
\usepackage{relsize}
\usepackage{graphicx}

\author{Dmitri Bykov$^{a,b}$\footnote{Email: dbykov@maths.tcd.ie}
 \\ $^a$ {\it School of Mathematics,
Trinity College, Dublin 2, Ireland}\\$^{b}$ {\it Steklov
Mathematical Institute, Moscow, Russia} }
\abstract{By direct calculation in classical theory we derive the central extension of the off-shell symmetry algebra for the string propagating in $\mathrm{AdS}_{4}\times \mathbb{CP}^{3}$. It turns out to be the same as in the case of the $\mathrm{AdS}_{5}\times S^{5}$ string. We also elaborate on the $\kappa$-symmetry gauge and explain, how it can be chosen in a way which does not break bosonic symmetries.}
\title{Off-shell symmetry algebra \\of the $\mathrm{AdS}_4 \times \mathbb{CP}^3$ superstring}
\preprint{{\scriptsize TCDMATH 09-11}}
\begin{document}
\newcommand{\ud}{\frac{1}{2}}
\newcommand{\dd}{\partial}
\newcommand{\tr}{{\rm tr}}
\newcommand{\Rr}{{\rm R}}
\newcommand{\Ll}{{\rm L}}
\newcommand{\Oo}{{\rm O}}
\newcommand{\Hh}{{\rm H}}
\newcommand{\Pp}{{\rm P}}
\newcommand{\Ee}{{\rm E}}
\newcommand{\Str}{\mathrm{Str}}
\onehalfspacing
\section{Introduction}

Relatively recently a new example of the $AdS/CFT$ correspondence \cite{M}, \cite{GKP}, \cite{W} was put forward --- the so-called ABJM model \cite{ABJM}. On the string theory side one deals with an $\mathrm{AdS}_{4}\times S^7 / Z_{k}$ near-horizon limit of a solution in 11-dimensional supergravity describing a stack of coincident M2-branes at a $Z_{k}$-orbifold singularity. The $Z_k$ acts on the $S^7$ in a peculiar way: namely, if one considers the Hopf fiber bundle $\pi: S^{7}\to \mathbb{CP}^3$ with fiber $S^1$, the $Z_k$ reduces the circumference of the circle by $k$ times, so in the limit $k\to \infty$ one gets rid of the circle completely, and we are left with the projective space $\mathbb{CP}^3$. The gauge theory dual to this $\mathrm{AdS}_4 \times \mathbb{CP}^3$ background is the $N=6$ supersymmetric Chern-Simons theory in three space-time dimensions (supersymmetry implies that the theory contains matter fields and is not topological for this reason).

Quite similar to the $\mathrm{AdS}_{5}\times S^{5}$ model, various signs of integrability have been discovered in this case, too. Namely, on the gauge theory side, integrability of the two-loop Hamiltonian (the one-loop Hamiltonian vanishes due to a discrete symmetry) was found in \cite{MZ} by direct check. Soon after this the algebraic curve for corresponding classical solutions and the all-loop asymptotic Bethe-ansatz were proposed \cite{GV1}, \cite{GV2}. Under the assumption of $su(2|2) \oplus u(1)$ symmetry algebra, the exact factorizable S-matrix was found in \cite{AN}. In the same paper the authors diagonalized the S-matrix and derived the Bethe ansatz equations, which agreed with those of \cite{GV2}.

On the string theory side, from the fact that the target space under consideration is maximally (super)symmetric, it follows that the string sigma model can be formulated as a coset model \cite{AF}. Using the coset formulation, one finds a Lax representation \cite{AF}, from which classical integrability follows.

However, the issue of integrability in the $\mathrm{AdS}_{4}\times \mathbb{CP}^{3}$ model has not been fully resolved so far. First of all, in a string theory calculation of one-loop correction to the spinning string (that is, a string with two charges $J$ and $S$) energy a mismatch was found with the Bethe ansatz prediction \cite{AAB}. Subsequently this result was confirmed by a calculation of the energy correction to a different string configuration --- the so-called circular string, which is a rational classical solution of the sigma-model \cite{TsM}. There has not been any convincing explanation of the mismatch so far. One of the explanations relies on the possible modification of $h(\lambda)$ (effective string tension, or coupling constant) due to loop corrections. Since $h(\lambda)$ enters the dispersion relation of the giant magnon, which in turn can be derived from the centrally-extended supersymmetry algebra as a BPS (multiplet-shortening) condition, the calculation of loop corrections to the central extension would prove useful and could help finally settle the issue. 

Another puzzle in the integrability program is that of the so-called 'heavy modes' in the BMN expansion on the string theory side. Indeed, the quadratic (leading) order of the BMN expansion was found in \cite{NT} and confirmed in \cite{AF}, and it follows from these papers that, apart from a multiplet of light particles of mass $\frac{1}{2}$ there's also a multiplet of heavy particles of mass $1$. The heavy excitations are not among the elementary excitations of the spin chain, namely, they contain two elementary momentum-carrying Bethe roots, which suggests that they're some sort of 'bound pair' of elementary magnons.

A possible resolution of this problem has been recently put forward in \cite{Zarembo}. The idea is that loop corrections remove the heavy particles from the spectrum. Namely, the pole of the corresponding heavy-particle propagator disappears, once, say, a one-loop correction is taken into account. This is related to the fact that the mass of the heavy particles lies precisely at the two-particle production threshold of light-particles and, of course, on the interactions of the theory, which are non-relativistic, since we move away from the strict BMN limit.

The central extension of the supersymmetry algebra in the $AdS_5 \times S^5$ case was introduced in \cite{B}. If the symmetry algebra of the $AdS_4 \times \mathbb{CP}^3$ superstring were altered as compared to the $\mathrm{AdS}_{5}\times S^{5}$ case, this could perhaps give some clues to the solution of the massive modes problem. However, as we explain below, the central extension is the same.

Other aspects of integrability of the $\mathrm{AdS}_{4}\times \mathbb{CP}^{3}$ have been studied \cite{Sundin}, namely near-BMN corrections to the energies of states in certain sectors were calculated therein.

The paper is organized as follows. In section 2 we give a definition of the quotient (coset) space that we are going to use. In section 3, we proceed to impose the light-cone gauge. Next, in section 4 we discuss the transformation properties of all the physical fields of the sigma-model under the residual light-cone global symmetry group. In section 5 we describe the kappa-symmetry gauge, which respects the global bosonic symmetries of the light-cone gauge. In section 6, we derive the central extension through the calculation of Poisson brackets. In carrying out the calculation we closely follow \cite{AFZ}, for instance, we use the so-called "hybrid" expansion introduced therein. In the Appendix the reader will find the explicit form of the necessary matrices, all global charges written out in terms of the fields, the Poisson brackets of these fields, as well as a general discussion of geodesics in $\mathbb{CP}^{3}$.

\section{Quotient space}

The $\mathrm{AdS}_4 \times \mathbb{CP}^3$ background (which we denote by $\mathcal{M}$ in what follows) is a ten-dimensional manifold, which admits the action of a topological group $\mathrm{G}=\mathrm{OSP}(6 | 2,2)$. The latter is a supergroup, which has $\mathrm{O}(6)\times \mathrm{USP}(2,2)$ as its maximal bosonic subgroup. The supergroup acts transitively on the manifold, the stabilizer of an arbitrary point $x_0$ in $\mathcal{M}$ being $\mathrm{H}=\mathrm{U}(3) \times \mathrm{O}(3,1)$. Thus, $\mathcal{M}$ is homeomorphic to $\mathrm{G} / \mathrm{H}$, the latter equipped with quotient topology.

Action of group $\mathrm{G}$ on the manifold $\mathcal{M}$ means that for a point $x_0$ in $\mathcal{M}$ and $g$ in $\mathrm{G}$ corresponds another point $x_1\equiv g( x_0 )$, and this correspondence is compatible with the group structure. Below we find this transformation law in suitable coordinates on $\mathcal{M}$.

$\mathbb{CP}^3$ may be viewed as the space of orthogonal complex structures in $\mathbb{R}^6$. Indeed, $U(3)\subset O(6)$ is the subgroup preserving a given complex structure, which we denote $K_6$ and, following \cite{AF}, choose in the form $K_6 = I_3 \otimes i \sigma_2$ ($I_3$ is the $3\times 3$ identity matrix). Then the Lie subalgebra $u(3)\subset o(6)$ is described by $6\times 6$ matrices, commuting with $K_6$. In other words, as vector spaces, $o(6)=u(3)\oplus V_\perp$.

The quotient vector space $W$, which describes the tangent space $T_x \mathcal{M}$ (tangent spaces are isomorphic for all $x$, since $\mathcal{M}$ is a manifold), is described by skew-symmetric matrices (elements of $o(6)$, that is) which anticommute with the complex structure. Indeed, we notice that for any $\omega \in O(6)$ the adjoint action $\omega K_6 \omega^{-1}$ is again a complex structure. For $\omega$ sufficiently close to unity $\omega=1+\epsilon + ...$, thus, $(K_6+[\epsilon, K_6])^2 +O(\epsilon^2) = -I_6$. Linear order in $\epsilon$ gives $\{K_6, [\epsilon, K_6]\}=0$. Define a map $f: o(6) \to o(6)$ by $f(a)=[a,K_6]$. Since ${\rm Ker}(f)={\rm u}(3)$, $W$ is isomorphic to ${\rm Im}(f)$. 
One can also check that if $g(b)\equiv\{K_6,b\}=0$, then $b\in{\rm Im}(f)=W$. \footnote{In fact, this choice of representatives in the quotient space becomes canonical once we adopt the Killing scalar product (since $f$ is skew-symmetric with respect to this scalar product $\tr(a,f(c))=-\tr(f(a),c)$). Indeed, for $a\in u(3)$ and $b\in {\rm Im}(f)$ we have $\tr (ab) = \tr(a[c,K_6])=\tr(a c K_6-a K_6 c)=0$, since $[a,K_6]=0$). This justifies the use of the symbol $V_\perp$ for $W$.} Let us note in passing that all of the above can be summarized by the following exact sequence of vector space homomorphisms ($i$ being inclusion):
\begin{equation}\label{exactseq}
0\to u(3)\overset{i}{\to} o(6) \overset{f}{\to} o(6) \overset{g}{\to} \mathbb{R}^{N},
\end{equation}
$\mathbb{R}^{N}$ being the vector space of symmetric matrices.

 It is easy to construct a basis in this linear space explicitly. Denoting by $J_1, J_2, J_3$ the three generators of $O(3)$ in the vector $3$ representation (see Appendix for an explicit form), we get: 
\begin{equation}
V_\perp = \textrm{Span} \{ J_i \otimes \sigma_1 ; J_i \otimes \sigma_3 \}
\end{equation}
To make contact with the notations of \cite{AF} we will write out the $T_i$ generators used in their paper in terms of the basis introduced above:
\begin{equation}
T_{1,3,5}=  J_{1,2,3} \otimes \sigma_3,\; T_{2,4,6}= J_{1,2,3} \otimes \sigma_1 .
\end{equation}
The main property which these generators exhibit and which will be important for us is the following:
\begin{equation}
\{T_1, T_2\}=\{T_3, T_4\}=\{T_5, T_6\}=0.
\end{equation}
For the following it is convenient to introduce the complex combinations
\begin{equation}
\mathcal{T}_1 = \frac{1}{2} (T_1 - i T_2),\;\;\mathcal{T}_2=\frac{1}{2} (T_3 - i T_4).
\end{equation}
$\bar{\mathcal{T}}_1$ and $\bar{\mathcal{T}}_2$ will denote the conjugate combinations.

\section{Light-cone gauge}

An extensive review of the light-cone gauge quantization of the $\mathrm{AdS}_{5}\times S^{5}$ superstring (which can be generalized to other maximally symmetric spaces), among many other things, can be found in the review \cite{review}. We introduce the light-cone coordinates:
\begin{equation}
x_{+}=\frac{1}{2}(\varphi + t),\;\; x_{-}=\varphi -t
\end{equation}
The corresponding canonical momenta $p_+$ and $p_-$ are conjugate to $x_-$ and $x_+$ respectively. Recall that the light-cone gauge comprises two conditions: $x_+ = \tau,\;p_{+}=\textrm{const}.$ We would like our string Lagrangian (and, consequently, Hamiltonian) not to depend on time $\tau$ even after the light-cone gauge is imposed. This requirement leads us to the following choice of parameterization for the coset element:
\begin{equation}\label{coset}
g= g_O g_\chi g_B,
\end{equation}
where $g_O = \exp{\left(\frac{i}{2}t \Gamma_0 + \frac{\varphi}{2} T_6\right)}$, $g_\chi=\exp{\chi}$, $g_B=\exp{\left(\frac{\alpha}{2} T_5\right)} \,g_{\rm{\mathbb{CP}}}\,g_{{\rm AdS}}$. We have chosen the coset representative $g_{\mathrm{AdS}}$ for AdS space in a way similar to the one in \cite{FPZ}:
\begin{equation}
g_{\mathrm{AdS}}=\frac{1}{\sqrt{1+\frac{z^2}{4}}}\, \left(1+\frac{i}{2} \sum\limits_{i=1}^{3} z_i \Gamma_i\right),
\end{equation}
where $z^2 = -\sum\limits_{i=1}^{3} z_i^2$. The matrix $g_{{\rm \mathbb{CP}}}$ gives, in turn, a parametrization of $\mathbb{CP}^{2}$ and is an obvious reduction of the coset element from \cite{AF}:
\begin{equation}
g_{{\rm \mathbb{CP}}}=I+\frac{1}{\sqrt{1+|w|^2}} \left(W+\bar{W}\right)+\frac{\sqrt{1+|w|^2}-1}{|\omega|^{2} \sqrt{1+|w|^{2}}} \left(W \bar{W}+\bar{W} W\right),
\end{equation}
where
\begin{equation}\label{Wdef}
W = w_1 \mathcal{T}_1 + w_2 \mathcal{T}_2,\;\;\bar{W}=\bar{w}_1 \bar{\mathcal{T}}_1+\bar{w}_{2} \bar{\mathcal{T}}_2.
\end{equation}

$\chi$ is the fermionic matrix of the following form:
$$\chi=\begin{bmatrix}
0&\theta\\
\eta&0
\end{bmatrix},\;
\theta=\begin{bmatrix}
n_{11}&\cdots &n_{16}\\
\vdots& &\vdots\\
n_{41}&\cdots&n_{46}\\
\end{bmatrix},
\;\eta=-\theta^T C_4 .$$
The reality condition of the algebra also ensures that the  two lower lines of $\theta$ are complex conjugates of the upper lines (we refer the interested reader to \cite{AF} for more information regarding this and other properties of the coset):
\begin{equation}\label{cc}
n_{3j} = - n_{2j}^*,\;n_{4j}=n_{1j}^* . 
\end{equation}

It is now easy to see that with this choice the current $A\equiv g^{-1}dg$, out of which the Lagrangian is built, does not explicitly depend on world-sheet time $\tau$. To make this property even more obvious, we rewrite the first exponent $g_O$ in terms of the light-cone coordinates:
\begin{equation}
g_O=\exp{\left(\frac{i}{2}x_+ \Sigma_+ + \frac{i}{4} x_- \Sigma_-\right)},
\end{equation}
where we have introduced $\Sigma_\pm = \pm \Gamma_0 - iT_6 = {\rm diag} \{\pm \Gamma_0 ; -i T_6\}$.

As is usual for gauge fixing procedures, after fixing the gauge we lose a certain amount of symmetry. The next problem we are going to tackle is to define the symmetry subgroup of $G$ which is left after imposition of the light-cone gauge. The subgroup of such transformations will be denoted by $G_{lc}$. Its Lie algebra consists of matrices which commute with the light-cone direction $\Sigma_+$. The block-diagonal bosonic subalgebra $g_{lc}^{Bose}$ is furnished  by matrices which commute with both $\Gamma_0$ and $T_6$ (i.e. the precise combination $\Sigma_+$ is only important for the fermionic part of the algebra, that is, for the supercharges). One can explicitly check that $g_{lc}^{Bose}=\mathrm{su}(2)\oplus \mathrm{su}(2) \oplus \mathrm{u}(1)$. One of these $\mathrm{su}(2)$s comes from the requirement of commutation with $\Gamma_0$, whereas $\mathrm{su}(2)\oplus \mathrm{u}(1)$ is the subalgebra of matrices, which commute with $T_6$. One can vaguely refer to the former as the $\mathrm{su}(2)$ coming from ${\rm AdS}_4$ whereas the latter is the algebra originating from $\mathbb{CP}^3$. Schematically the position of the embeddings of the corresponding matrices looks as follows:
\begin{equation}
\omega=\begin{pmatrix}
su(2)|_{4\times 4}^{\mathrm{AdS}}&0 & 0\\
0& u(1)|_{2\times 2}^{\mathbb{CP}}&0\\
0&0&su(2)|_{4\times 4}^{\mathbb{CP}}\\
\end{pmatrix}
\end{equation}
For a precise description of these matrices see Appendix.

Suppose we now want to calculate the full algebra $g_{lc}$, which is left after the light-cone condition has been imposed. This means, that we will include supersymmetry transformations, and will no longer limit ourselves to the bosonic part $g_{lc}^{Bose}$. Then, as one can explicitly check, the full algebra turns out to be $g_{lc}={\rm su}(2|2)\oplus {\rm u}(1)$. It is precisely this algebra that acquires a central extension after quantization. We leave a more elaborate discussion of this point until section 6.

\section{Transformation properties of the fields}

\subsection{Bosons}

In this section we will find out the transformation properties of the fields, both bosonic and fermionic, under $G_{lc}^{Bose}$ (or $g_{lc}^{Bose}$ in infinitesimal form). It is important to notice that $G_{lc}^{Bose} \subset H$. Let us act on the coset element (\ref{coset}) from the left by a bosonic group element from $G_{lc}^{Bose}$, which we denote by $\exp{a}$, assuming that $a$ is in the Lie algebra $g_{lc}$:
\begin{equation}\label{trans}
g \to e^a g = g_O e^{ad_a} (g_\chi ) e^{ad_a} (g_B ) e^a ,
\end{equation}
where we have taken into account that $[a,\Gamma_0]=[a,T_6]=0$. The exponent at the very right is irrelevant, since it belongs to the stabilizer, and, as such, does not change the corresponding conjugacy class. The above formula shows, then, that the bosons and fermions are in the adjoint representation of $G_{lc}^{Bose}$. To be absolutely clear, we will describe the transformation properties of the fields even more explicitly. We work in the basis of $\gamma$-matrices described in the appendix, from which it follows that for $k=1,2,3$ we have $\gamma_{k}=i\sigma_2 \otimes \sigma_{k}$, so that the $\mathrm{AdS}$ part of the coset is written in the following form:
\begin{equation}
g_{\mathrm{AdS}}= \frac{1}{\sqrt{1+\frac{z^2}{4}}}\, 
 \begin{pmatrix}
I_2 &\sum\limits_{i=1}^{3} z_{i}\sigma_{i }\\
-\sum\limits_{i=1}^{3} z_{i}\sigma_{i }&I_{2}
\end{pmatrix}
\end{equation}
As described above, under the action of the $SU(2)$ group from $\mathrm{AdS}$ this element transforms in the adjoint. Thus, introducing notation $Z\equiv \sum\limits_{i=1}^{3} z_{i}\sigma_{i } $, we get ($\Delta$ is the diagonal embedding defined in Appendix):
\begin{equation}
g_{\mathrm{AdS}} \to \Delta(\omega) g_{\mathrm{AdS}} \Delta(\omega^{\dagger})= \frac{1}{\sqrt{1+\frac{z^2}{4}}}\, 
 \begin{pmatrix}
I_2 &\omega Z \omega^{\dagger}\\
-\omega Z  \omega^{\dagger}&I_{2}
\end{pmatrix}
\end{equation}
Since $Z$ is a traceless Hermitian matrix, $Z\to \omega Z \omega^{\dagger}$ with $\omega \in SU(2)$ defines a vector representation. In order to single out other irreducible representations, we introduce three complex combinations of the $T_i$ generators:
\begin{equation}
\tau_1 = T_1+i T_2 ;\; \tau_2 = T_3 + i T_4 ;\; \tau_3 = T_5 + i T_6 .
\end{equation}
Then we can rewrite $\sum\limits_{i=1}^{4} \beta_i T_i = \beta_1^+ \tau_1 + \beta_2^+ \tau_2 + {\rm C.c.}$, where $\beta_{1,2}^+$ are a set of new (complex) coordinates. It is now easy to check that under $\mathrm{su}(2)\oplus \mathrm{u}(1)$ from $\mathbb{CP}^3$ these coordinates form a complex doublet and, consequently, $\beta_i$ transform as $2^1 + 2^{-1}$, where the exponent refers to $\mathrm{u}(1)$ charge. Indeed, let us denote the ${\rm su}(2)$s from $\mathrm{AdS}_4$ and $\mathbb{CP}^3$ as ${\rm su}(2)_R$ and ${\rm su}(2)_L$ respectively. Let
\begin{equation}
\omega = \begin{pmatrix}
\omega_{1}&0 \\
0& \widetilde{\omega}\\
\end{pmatrix} \;\mbox{with}\; \widetilde{\omega} =  \begin{pmatrix}
\omega_{u}&0 \\ 0& \omega_2 \end{pmatrix}
\end{equation}
be a generic transformation matrix. Using the fact that for compact groups the exponential map is surjective, we introduce an explicit parameterization for these matrices: $\omega_{u}=\exp{(-\alpha \,i\sigma_{2})}$ and $\omega_{2}=\exp{(\sum\limits_{i=1}^{3} \, \delta_{i}\,s_{i})}$. Now, the non-zero part of the top line of the matrix $W$ (see Appendix) can be written in the form
\begin{equation}
\widehat{\omega}= (\omega_{1}, \omega_{2})\otimes \frac{1}{2}(1,-i).
\end{equation}
Acting on it by $\omega_{2}^{-1}$ from the right, we obtain the transformation law:
\begin{equation}\label{omtrans}
(\omega_{1}, \omega_{2}) \to (\omega_{1}, \omega_{2}) \, \left( \exp{(\pm i\sum\limits_{j=1}^{3} \delta_{j} \sigma_{j})}\right)^{{\rm T}},
\end{equation}
which is the canonical $SU(2)$ action (defined on row-vectors, rather than on column-vectors).

One can actually propose an even stronger statement, namely that under the adjoint action of $H=U(3)$ the $\tau_i$ are in the $3$ irrep, that is, they transform as a complex triplet. This means that $T_i$ are in the $3+\bar{3}$ representation. One of the consequences of this fact is the following interesting property: those skew-symmetric $6\times 6$ matrices (that is, the ones in $\mathrm{so}(6)$) which commute with $T_6$ simultaneously commute with $T_5$. Thus, transformations which leave $T_6$ invariant also leave $T_5$ invariant.

\subsection{Fermions}

Next we turn to the transformation properties of the fermions $\chi$. They form a representation of ${\rm su}(2)\oplus {\rm su} (2)\oplus {\rm u}(1)$, and we need to decompose it into irreducibles. We will proceed in analogous way to what was done for the bosonic case. The fermionic matrix $\theta$ undergoes the following transformation
\begin{equation}
\theta \to \omega_{1} \, \theta \,\, \widetilde{\omega}^{-1}
\end{equation}
As is described in Appendix, the matrix basis of the Lie algebra of $su(2)|^{\mathbb{CP}}_{4\times 4}$ looks as $A\otimes B$, where $B$ is either the identity matrix $I_2$ or the skew-symmetric matrix $i\sigma_2$. Moreover, the $u(1)$ generator is simply $i\sigma_2$. Thus, it makes sense to single out the components of the fermionic matrix, which correspond to eigenvalues of the $\sigma_{2}$ matrix. In this way we obtain:
\begin{equation}
\theta = \theta^{(+1)}+\theta^{(-1)}+\theta^{(0)}_{+}+\theta^{(0)}_{-},
\end{equation}
where $\theta^{(\pm 1)}$ have non-zero columns 1 and 2, $\theta^{(0)}_{\pm}$ have zero columns 1 and 2. To simplify some expressions below we will for the moment cut off the zero columns from all of these matrices, namely, we will regard $\theta^{(\pm 1)}$ as $2\times 4$ matrix and $\theta^{(0)}_{\pm}$ as $4\times 4$ matrix. It should be clear from the context, if these matrices should be embedded into bigger ones. Then these matrices can be defined as follows:
\begin{eqnarray}
\theta^{(+1)}= \kappa^{(+1)} \otimes (1, -i),\; \theta^{(-1)}= \kappa^{(-1)} \otimes (1, i),\\
\theta^{(0)}_{+}= \chi^{+0} \otimes (1, -i), \; \theta^{(0)}_{-}=\chi^{-0} \otimes (1, i).
\end{eqnarray}
One can consult the Appendix for an explicit form of the matrix $\theta$ in terms of all of these components.

Being multiplied by $\omega_{u}^{-1}$ from the right, obviously $\chi^{\pm 0}$ remain unaltered, whereas $\kappa^{(\pm 1)}$ transform as follows:
\begin{equation}
\kappa^{(\pm 1)} \to e^{\pm i \alpha} \kappa^{(\pm 1)}.
\end{equation}
On the other hand, being multiplied by $\omega_{2}^{-1}$ from the right, $\kappa$ is unaltered, but $\chi^{(\pm 0)}$ transform as follows:
\begin{equation}
\chi^{(\pm 0)} \to \chi^{(\pm 0)} \left( \exp{(\pm i\sum\limits_{j=1}^{3} \delta_{j} \sigma_{j})}\right)^{{\rm T}},
\end{equation}
which is the same transformation law as (\ref{omtrans}). We have written out the components of the matrices $\kappa^{(+ 1)}$ and $\chi^{+ 0}$ explicitly in the Appendix (see (\ref{chi+})). In terms of these components the transformation properties of the fermions are as follows:
\begin{eqnarray}\label{transform}
\kappa^{a,+1} \to e^{i\phi} (\omega_{1})^{a}_{\;b}\, \kappa^{b, +1} ,\\ \nonumber
(\chi^{+0})^{a}_{\alpha} \to (\omega_{1})^{\; a}_{\;\; b}\, (\omega_{2})^{\;\beta}_{\;\;\alpha} (\chi^{+0})^{b}_{\beta}.
\end{eqnarray}
In other words, if regarded as a matrix, $\chi=\{\chi^{a}_{\alpha}\}$ transforms as $\chi \to \omega_{1} \chi (\omega_{2})^T$. Of course, we also need to know how combinations like $n_{11}-i n_{12}$ (which comprise $\kappa^{(-1)}$ and $\chi^{-0}$) transform. It turns out that they're in the conjugate representation with respect to the $\mathbb{CP}^{3}$ part of the algebra, and in the same representation of the $\mathrm{AdS}$ part of the algebra. It will be useful to give $\chi^{-}$ the transformation properties identical to those of  $\chi^{+}$ and to convert $\kappa^{-1}$ to the representation conjugate to the one of $\kappa^{+1}$. This is convenient, because $\chi^{\pm}$'s are not charged with respect to the $U(1)$, whereas $\kappa^{\pm}$ have opposite $U(1)$ charges. Since $\kappa$'s carry opposite $U(1)$ charges, it is also natural to give them opposite transformation properties with respect to the $SU(2)$, which comes from $\mathrm{AdS}$ (they're uncharged with respect to the $SU(2)$ which comes from $\mathbb{CP}^3$).

It is always possible to change the transformation properties of the fields in this fashion, since the fundamental and conjugate-fundamental representations of $SU(2)$ are equivalent, which means that there's a matrix $C\in SU(2)$, securing a relation
\begin{equation}
\omega^{*}=C \omega C^{-1} \,\mbox{for}\, \omega \in SU(2).
\end{equation}
In fact, $C=i\sigma_{2}$. Again, one can find the explicit form of the relevant combinations in (\ref{chi-}): they transform as in (\ref{transform}), apart from the fact that the exponent $e^{i\phi}$ needs to be replaced with $e^{-i\phi}$ to account for the opposite transformation property with respect to the $U(1)$.

The indices have been chosen to suggest, what the representations of the fermions are. One can summarize the transformation properties described above as follows: $\kappa_{1,2}^{\pm 1}$ are in the fundamental of ${\rm su}(2)_R$, the $\pm$ carrying opposite charges with respect to the ${\rm u}(1)$, whereas $\chi_{\alpha\dot{\beta}}^{\pm 0}$ are in the bifundamental of ${\rm su}(2)_R \oplus {\rm su}(2)_L$ and neutral under ${\rm u}(1)$. From (\ref{cc}) it follows that the two lower lines of the matrix $\theta$ transform in an analogous way. In total we have 12 complex fermion fields, which have been grouped into irreps as $\kappa^{\pm1}_{\alpha}$, $\chi_{\alpha\dot{\beta}}^{\pm 0}$.

\section{$\kappa$-symmetry gauge}

As is well-known, string sigma-models in the Green-Schwarz formulation possess, besides diffeomorphism and Weyl invariance, another sort of gauge invariance --- the $\kappa$-symmetry, which is fermionic in the sense that the gauge (=local) parameters are fermionic (denoted by $\epsilon$ in what follows). Existence of such transformations was first observed by Green and Schwarz for the flat background, however, it was also discovered for string models in other backgrounds, including the $\mathrm{AdS}_5 \times S^5$ case. It is of course a remarkable property that the same sort of invariance also holds in the $\mathrm{AdS}_4 \times \mathbb{CP}^3$ case under consideration \cite{AF}.

Once the existence of $\kappa$-symmetry is established, one needs to choose a gauge (a representative in every gauge orbit). This can be done in various ways, however, one aims at preserving as much global symmetry as possible during this process, since global symmetry allows for a better classification of field multiplets and ultimately leads to a simpler formulation of the theory. In our case this requirement means that the whole $G_{lc}^{Bose}$ should be preserved. Let us now elaborate on how this can be done. As found in \cite{AF}, the leading order in $\epsilon$ of the $\theta$ field variation is
\begin{equation}\label{kappa}
\delta \theta = \begin{bmatrix}
0&0&\epsilon_{1}&\epsilon_{2}&-i \epsilon_{2}&-i\epsilon_{1}\\
0&0&\epsilon_{3}&\epsilon_{4}&-i\epsilon_{4}&-i\epsilon_{3}\\
0&0&\star&\star&\star&\star\\
0&0&\star&\star&\star&\star
\end{bmatrix},
\end{equation}
stars denoting complex conjugated variables, totally parallel to (\ref{cc}). One might observe that (\ref{kappa}) is the upper right block of a generic fermionic matrix $\vartheta$, which has the property $f_{1}(\vartheta)\equiv[\vartheta, \Sigma_+]=0$ (not to mention reality conditions discussed numerous times above). Let $W_{F}$ be the full fermionic vector space. Factorizing over the gauge-equivalent combinations, we thus get $W_{F}/{\rm Ker} f_{1} \sim {\rm Im} \,f_{1}$. Let us check that ${\rm Im} \,f_{1}$ is invariant under the action of $G_{lc}^{Bose}$. If $a$ belongs to $g_{lc}^{Bose}$ (which means that $[a, \Sigma_+ ]=0$) and $c = [d, \Sigma_+ ] \in {\rm Im} f_{1}$, then $e^{{\rm ad}_a}(c)=[e^{{\rm ad}_a}(d),\Sigma_+ ] \in {\rm Im} \, f_{1}$.
Restricting the fermion to ${\rm Im}\, f_{1}$ corresponds to setting
\begin{equation}\label{gaugek}
n_{15}=i n_{14},\;n_{16}=i n_{13},\;n_{25}=i n_{24},\;n_{26}=i n_{23},
\end{equation}
which is the explicit form of the gauge we will be using in what follows. Then, as one can easily see from (\ref{chi+}), (\ref{chi-}), $\chi^{+0}=\chi^{-0}\equiv \chi^{0}$, so we effectively get rid of one of the multiplets. As a result, we are left with 8 complex fermions, which is the correct number for supersymmetry. We want to emphasize that this choice of kappa-symmetry gauge is equivalent to requiring that for any $\chi$ there's a matrix $\xi$ such that $\chi = [\Sigma_{+},\xi]$. Obviously, this matrix is not unique.

\section{Central extension}

In section 4 we discussed the representation of the fermionic fields under the action of the bosonic part of the symmetry algebra. The odd part of the symmetry algebra may be realized in a way very similar to the fermionic fields --- that is, as odd elements of a $4|6\times 4|6$ matrix. This means, that the representation of supercharges is a subrepresentation of the one the fermions transform under. Indeed, as compared to the fermions, there is an extra condition on the supercharges, namely, the requirement that they must commute with $\Sigma_+$. This leaves only four complex independent supercharges, as expected for an $su(2|2)$ algebra.

In the context of the $su(2)\oplus su(2)$ algebra we use Latin indices for the $\mathrm{AdS}$ part and Greek indices for the $\mathbb{CP}^3$ part. The generators of $\mathrm{su}(2|2)$ can be conveniently described by two traceless bosonic operator-valued matrices $R_\alpha^\beta$ and $L_a^b$, and an operator-valued (complex) fermionic matrix $\mathcal{Q}_\alpha^a$ . It should thus be clear that $L$ and $R$ describe $\mathrm{AdS}$ and $\mathbb{CP}^3$ rotations, respectively. With respect to the Poisson bracket, the entries of these matrices form the following Lie algebra:
\begin{eqnarray}\label{central}
[ \Rr_\alpha^\beta , \Rr_\gamma^\delta ] = \delta_\alpha^\delta \Rr_\gamma^\beta -  \delta_\gamma^\beta \Rr_\alpha^\delta \\ 
\nonumber [ \Ll_a^b , \Ll_c^d ] = \delta_a^d \Ll_c^b - \delta_c^b \Ll_a^d \\
\nonumber \{ \mathcal{Q}_\alpha^a , \bar{\mathcal{Q}}^\beta_b \} =  \delta_\alpha^\beta \Ll_b^a-\delta_b^a \Rr_\alpha^\beta + \frac{1}{2}\delta_b^a \delta_\alpha^\beta \Hh\\
\nonumber \{\mathcal{Q}_\alpha^a , \mathcal{Q}_\beta^b \} = \epsilon_{\alpha\beta} \epsilon^{ab} \Pp_1\\
\nonumber \{ \bar{\mathcal{Q}}^\alpha_a , \bar{\mathcal{Q}}^\beta_b \} = \epsilon_{ab} \epsilon^{\alpha\beta} \Pp_2 . 
\end{eqnarray}
Obviously, $P_{2}=\bar{P}_{1}$. Besides, one might check that the bosonic part (the first two lines) is precisely $\mathrm{su}(2)\oplus \mathrm{su}(2)$ if one makes the following identifications: $R_1^1=\frac{1}{2}\sigma_3,\; R_1^2=\sigma_- ,\;R_2^1=\sigma_+$.

Using Noether's theorem, one can find the matrix of supercharges, that is, a divergence-free vector field with values in the Lie algebra $osp(6|2,2)$:
\begin{equation}
J^\alpha = \; g \left( \gamma^{\alpha \beta} A_\beta^{(2)}+\frac{\kappa}{2} \epsilon^{\alpha\beta} (A_\beta^{(3)}-A_\beta^{(1)}) \right) g^{-1}
\end{equation}
Here, as usual, the upper indices in brackets denote the corresponding component of the current $A_\alpha$ under the $Z_4$ grading. The components of $J_\alpha$ under the decomposition over the Lie algebra basis are conserved currents corresponding to various charges, both bosonic and fermionic.

We proceed by imposing the light-cone gauge. To do that, we will use the first-order formalism, as described in \cite{FPZ}. This is not necessary, but simplifies the calculations. Thus, we rewrite the Lagrangian in the following form:
\begin{equation}\label{firstorder}
\frac{2\pi}{\sqrt{\lambda}}\mathcal{L}=\frac{1}{2 \gamma^{00}} \Str((\mathcal{P}_{0})^{2})-\Str(\mathcal{P}_{0}^{(2)}(A_{0}+\frac{\gamma^{01}}{\gamma^{00}} A_{1}))+\frac{1}{2 \gamma^{00}} \Str((A_{1}^{(2)})^{2}) - \frac{\kappa}{2} \epsilon^{\alpha\beta} \Str(A^{(1)}_{\alpha} A^{(3)}_{\beta}).
\end{equation}
In the first term we could have written $\mathcal{P}_{0}^{(2)}$, but all other terms decouple anyway, so they may only contribute to the normalization of the path integral, which is irrelevant so far. In fact, the physical meaning of $\mathcal{P}_{0}$ is that it provides for a decomposition of the momentum over the local (super)vielbein (at least when the Wess-Zumino term is neglected). Indeed, denoting by $X_{\mu}$ the set of all possible fields, $A_{0}^{(2)}\equiv-E_{\mu}^{a} \dot{X}_{\mu} T_{a}$, so, if one neglects the Wess-Zumino term, $p_{X_{\mu}}=E_{\mu}^{a} \, \Str(\mathcal{P}_0 T_{a})=E_{\mu}^{a} \,\mathcal{P}_{0,a}$. Thus, in this way we effectively avoid the complicated contributions to the explicit expressions of momenta, which come from the vielbein. The possibility of dropping the Wess-Zumino term in our case is justified by the fact that it does not contribute to any variables entering the algebra in the leading order. Indeed, for the calculation of the algebra we need the term in the supercharges, linear in the fermions, and the term in the bosonic charges, quadratic in the bosons, whereas the Wess-Zumino term is (at least) quadratic in the fermions.

From (\ref{firstorder}) one immediately reads off the Virasoro conditions:
\begin{eqnarray}
V_{1} &\equiv& \Str(\mathcal{P}_{0}^{(2)}A_{1})=0,\\
V_{2}&\equiv& \Str((\mathcal{P}_{0}^{(2)})^2 +(A_{1}^{2})^2)=0
\end{eqnarray}

It is important to note that, once the kappa-gauge has been imposed, the action of supersymmetry transformations on physical fields is given by $g\to e^{\epsilon} g e^{\widetilde{\epsilon}}$, where $\widetilde{\epsilon}$ is a compensating kappa transformation (and it is uniquely determined by $\epsilon$). This is in contrast to the action $g\to e^{\epsilon} g$, which (as described above) one has before imposing the kappa-symmetry gauge. This is not special to the case under consideration, but rather is a general property of superstring theories --- for instance, it is also present in the flat case \cite{Green}. This is very similar, for instance, to the Wess-Zumino gauge in supersymmetric theories: it manifestly breaks supersymmetry, but there's a symmetry of the gauge-fixed action, which is a combination of the supersymmetry transformation and a gauge transformation \cite{dWF}. 

To perform the calculation of the Poisson bracket we extensively use the formulas, obtained in Appendix. A straightforward calculation gives the following result for the central extensions entering formulas  (\ref{central}):
\begin{eqnarray}
P_{1}=-\frac{i}{2} \int\,d\sigma\, e^{-i x_{-}} \, x_{-}' = \frac{1}{2} e^{-i x_{-}(-\infty)} (e^{-ip}-1)=\frac{\xi}{2} (e^{-ip}-1),\\
P_{2}=\frac{i}{2} \int\,d\sigma\, e^{i x_{-}} \, x_{-}' = \frac{1}{2} e^{i x_{-}(-\infty)} (e^{ip}-1)=\frac{\xi^{*}}{2} (e^{ip}-1),
\end{eqnarray}
where we have introduced $\xi \equiv e^{-i x_{-}(-\infty)}$.

It is interesting to mention, that this sort of algebra may well be called worldsheet supersymmetry algebra, since it includes worldsheet charges $p$ and $H$, as well as the supersymmetry generators and other target-space charges. In fact, the only difference from the usual supersymmetry algebra is the fact that $p$ and $H$ are central. However, in the ordinary supersymmetry algebra, once one omits the Lorentz generators, the corresponding energy-momentum becomes central, too. Of course, there's no Lorentz algebra in the light-cone worldsheet theory, since it is not Lorentz-invariant. Even if it were, Lorentz symmetry is quite simple in two dimensions. Nevertherless, in this case there's a much more interesting counterpart, namely the $SL(2)$ group of outer automorphisms of the algebra \cite{Beisert}. It acts as the three-dimensional rotation group (or Lorentz group) \cite{HM}. These automorphisms are outer, since they do not preserve the reality properties of the fermionic charges.

\section{Conclusion}

In the first part of this paper we proposed a kappa-symmetry gauge, compatible with the bosonic $su(2)\oplus su(2) \oplus u(1)$ symmetries. The second part was devoted to the classical calculation of the central extension to the supersymmetry algebra in the framework of the so-called hybrid expansion. Calculation of the corresponding Poisson brackets between the supercharges led to the same result, as had been previously obtained for the $\mathrm{AdS}_{5}\times S^{5 }$ case. As a slight deviation from the main line of the text, in the appendix we present a general scheme for the analysis of geodesics in $\mathbb{CP}^{3}$ (which is of course also suitable for any other symmetric space).

\section*{Acknowledgments}

I am grateful to Sergey Frolov for suggesting that I work on the problem of off-shell algebra and for many useful and illuminating discussions in the course of work. I want to thank Ryo Suzuki for an interesting discussion about the geodesics in $\mathbb{CP}^3$ and Per Sundin for a number of interesting discussions. I also want to thank Gleb Arutyunov and Sergey Frolov for carefully reading the manuscript and contributing to its improvement by valuable comments. The manuscript was finalized while I was attending the school "Algebraic and combinatorial structures in quantum field theory" in Cargese, and I'm grateful to the organizers for a very stimulating atmosphere. My work was supported by the Irish Research Council for Science, Education and Technology, in part by grant of RFBR 08-01-00281-a and in part by grant for the Support of Leading Scientific Schools of Russia NSh-795.2008.1.

\appendix

\section{Matrices and all that}

The representation of $\gamma$ matrices used throughout the paper is as follows:
\begin{equation}\small
\gamma_{0}= \begin{pmatrix}
1 &0&0&0\\
0&1&0&0\\
0&0&-1&0\\
0&0&0&-1
\end{pmatrix},\;
\gamma_{1}= \begin{pmatrix}
0&0&0&1\\
0&0&1&0\\
0&-1&0&0\\
-1&0&0&0
\end{pmatrix},\;
\gamma_{2}= \begin{pmatrix}
0&0&0&-i\\
0&0&i&0\\
0&i&0&0\\
-i&0&0&0
\end{pmatrix},
\gamma_{3}= \begin{pmatrix}
0 &0&1&0\\
0&0&0&-1\\
-1&0&0&0\\
0&1&0&0
\end{pmatrix},\;
\end{equation}
One can observe that, as stated in the paper, for $k=1,2,3$ we have $\gamma_{k}=i\sigma_{2} \otimes \sigma_{k}$. Another matrix encountered in the text is
\begin{equation}
C_{4}=i\Gamma_{0}\Gamma_{2}=\begin{pmatrix}
0 &0&0&1\\
0&0&-1&0\\
0&1&0&0\\
-1&0&0&0
\end{pmatrix}
\end{equation}

The $J$-matrices --- generators of $SO(3)$ --- used in section 2, are defined as follows:
\begin{equation}
J_{1}=\begin{pmatrix}
0 &1&0\\
-1&0&0\\
0&0&0\\
\end{pmatrix}
,\;J_{2}=\begin{pmatrix}
0 &0&1\\
0&0&0\\
-1&0&0\\
\end{pmatrix},\;J_{3}=\begin{pmatrix}
0 &0&0\\
0&0&1\\
0&-1&0\\
\end{pmatrix}
\end{equation}
They satisfy the usual condition $[J_{k},\, J_{l}]=-\epsilon_{klm} J_{m}$.

The matrix $W$ defined in (\ref{Wdef}) looks as follows:
\begin{equation}\small
W=\left(
\begin{array}{llllll}
 0 & 0 & \frac{\omega_1}{2} & -\frac{i \omega_1}{2} & \frac{\omega_2}{2} & -\frac{i \omega_2}{2} \\
 0 & 0 & -\frac{i \omega_1}{2} & -\frac{\omega_1}{2} & -\frac{i \omega_2}{2} & -\frac{\omega_2}{2} \\
 -\frac{\omega_1}{2} & \frac{i \omega_1}{2} & 0 & 0 & 0 & 0 \\
 \frac{i \omega_1}{2} & \frac{\omega_1}{2} & 0 & 0 & 0 & 0 \\
 -\frac{\omega_2}{2} & \frac{i \omega_2}{2} & 0 & 0 & 0 & 0 \\
 \frac{i \omega_2}{2} & \frac{\omega_2}{2} & 0 & 0 & 0 & 0
\end{array}
\right)
\end{equation}
In the main text we have used a separate notation for the first row of this matrix:
\begin{equation}\small
\widehat{w}= (\frac{\omega_1}{2}, \;-\frac{i \omega_1}{2},\; \frac{\omega_2}{2},\; -\frac{i \omega_2}{2}) = (\omega_{1},\;\omega_{2})\otimes \frac{1}{2} (1,-i).
\end{equation}

The fermionic matrix $\theta$ looks as follows:
\begin{equation}\scriptsize\nonumber
\theta = 
\left(
\begin{array}{llllll}
 \kappa_{1}^{+1}-\kappa _{2}^{-1} & -i \left(\kappa _{1}^{+1}+\kappa _{2}^{-1}\right) & \frac{1}{2} \left(\chi^{+}_{1,1}-\chi^{-}_{1,2}\right) & -\frac{1}{2} i
   \left(\chi^{+}_{1,1}+\chi^{-}_{1,2}\right) & \frac{1}{2} \left(\chi^{+}_{1,2}+\chi^{-}_{1,1}\right) & -\frac{1}{2} i \left(\chi^{+}_{1,2}-\chi^{-}_{1,1}\right) \\
 \kappa _{1}^{-1}+\kappa _{2}^{+1} & i \left(\kappa _{1}^{-1}-\kappa _{2}^{+1}\right) & \frac{1}{2} \left(\chi^{+}_{2,1}-\chi^{-}_{2,2}\right) & -\frac{1}{2} i
   \left(\chi^{+}_{2,1}+\chi^{-}_{2,2}\right) & \frac{1}{2} \left(\chi^{+}_{2,2}+\chi^{-}_{2,1}\right) & -\frac{1}{2} i \left(\chi^{+}_{2,2}-\chi^{-}_{2,1}\right) \\
 -\bar{\kappa }_{1}^{-1}-\bar{\kappa }_{2}^{+1} & i \left(\bar{\kappa }_{1}^{-1}-\bar{\kappa }_{2}^{+1}\right) & \frac{1}{2}
   \left(\bar{\chi}^{-}_{2,2}-\bar{\chi}^{+}_{2,1}\right) & -\frac{1}{2} i \left(\bar{\chi}^{+}_{2,1}+\bar{\chi}^{-}_{2,2}\right) & \frac{1}{2}
   \left(-\bar{\chi}^{+}_{2,2}-\bar{\chi}^{-}_{2,1}\right) & -\frac{1}{2} i \left(\bar{\chi}^{+}_{2,2}-\bar{\chi}^{-}_{2,1}\right) \\
 \bar{\kappa }_{1}^{+1}-\bar{\kappa }_{2}^{-1} & i \left(\bar{\kappa }_{1}^{+1}+\bar{\kappa }_{2}^{-1}\right) & \frac{1}{2}
   \left(\bar{\chi}^{+}_{1,1}-\bar{\chi}^{-}_{1,2}\right) & \frac{1}{2} i \left(\bar{\chi}^{+}_{1,1}+\bar{\chi}^{-}_{1,2}\right) & \frac{1}{2}
   \left(\bar{\chi}^{+}_{1,2}+\bar{\chi}^{-}_{1,1}\right) & \frac{1}{2} i \left(\bar{\chi}^{+}_{1,2}-\bar{\chi}^{-}_{1,1}\right)
\end{array}
\right)\end{equation}
At this point we would like to remind the reader that the kappa-gauge corresponds to setting $\chi^{+}=\chi^{-}\equiv\chi$ in the matrix written above.

We can equally express the fields $\kappa^{\pm 1},\,\chi^{\pm 0}$ in terms of the $n_{ij}$, i.e. elements of the matrix $\theta$. Namely,
\begin{eqnarray}\label{chi+}
&\kappa_1^{+ 1} = \frac{1}{2} (n_{11} + i \,n_{12}),\;\kappa_2^{+ 1} = \frac{1}{2}(n_{21} + i \,n_{22})&\\
\nonumber
&\chi_{1\dot{1}}^{+ 0}= n_{13} + i \,n_{14},\;\chi_{1\dot{2}}^{+ 0}= n_{15} + i \,n_{16}&
\\ \nonumber
&\chi_{2\dot{1}}^{+ 0}= n_{23} + i \,n_{24},\;\chi_{2\dot{2}}^{+ 0}= n_{25} + i \,n_{26}&
\end{eqnarray}
The 'conjugate' combinations are
\begin{eqnarray}\label{chi-}
&\kappa^{-1}_{1}=\frac{1}{2}(n_{21} - i \,n_{22}),\;\kappa^{-1}_{2}=-\frac{1}{2} (n_{11} - i \,n_{12})&\\ \nonumber
&\chi^{-0}_{1\dot{1}}=n_{15} - i \,n_{16},\;\chi^{-0}_{1\dot{2}}= -(n_{13} - i \,n_{14})&\\ \nonumber
&\chi^{-0}_{2\dot{1}}=n_{25} - i \,n_{26},\;\chi^{-0}_{2\dot{2}}=-(n_{23} - i \,n_{24})&
\end{eqnarray}

First of all, we describe explicitly the matrix generators of the $su(2)\oplus su(2)\oplus u(1)$ algebra. We introduce the following matrices:
\begin{eqnarray}\nonumber 
&t_{k}=-\frac{i}{2} \Delta(\sigma_{k}), \; u=\begin{pmatrix}\; 0& 1\\
-1&0\\
\end{pmatrix}=i\sigma_{2},&\\ &s_{1}=\frac{1}{2} \sigma_{1}\otimes i\sigma_{2},\; s_{2}=-\frac{1}{2} i\sigma_{2}\otimes I_2 ,\; s_{3}= \frac{1}{2} \sigma_3 \otimes i\sigma_2 ,&
\end{eqnarray}
$I_2$ being the $2\times 2$ identity matrix and $\Delta$ the diagonal embedding: $\Delta(a)=I_2 \otimes a $. In these notations $t_{k}$ describe the $su(2)|_{4\times 4}^{\mathrm{AdS}}$, $u$ is the $u(1)|_{2\times 2}$ U(1)-charge from $\mathbb{CP}^3$ and $s_{k}$ describe the $su(2)|_{4\times 4}^{\mathbb{CP}}$. These matrices (after corresponding embeddings into $10\times 10$ matrices) satisfy the necessary reality conditions (for example, the $s_i$ are real) and the following commutation relations:
\begin{equation}
[t_{i}, \,t_{j}]=\epsilon_{ijk}\,t_{k},\;[s_{i}, \,s_{j}]=\epsilon_{ijk}\,s_{k},\;[t_{i},\,s_{j}]=[t_{i},\,u]=[s_{i},\,u]=0.
\end{equation}

\section{The charges}

In this appendix we write down the explicit expressions (in terms of the fields of the model) for all charges appearing in the symmetry algebra. We also present the Poisson brackets of the fields, so that it is easy to check that the charges do indeed satisfy the claimed symmetry algebra.

First of all, we find it necessary to explain the notation used below. More precisely, we need to explain how indices of various fields are lowered and raised, since without this understanding it is impossible to check the covariance of the expressions that we obtain, even if lower indices are always contracted with upper ones. 
First of all, $Z\equiv z_{i} \sigma_{i}$ and $P_{z} \equiv P_{z_{i}} \sigma_{i}$. For matrix elements of these matrices we use the notation $Z^{a}_{\;\; b}$ and $(P_{z})^{a}_{\;\;b}$ respectively. We need to use this shifted notation for the indices, since otherwise it would not be clear, what $Z_{1}^{2}$ and $Z_{2}^{1}$ stand for. The indices in our notation should be (as usual) read from left to right, that is, for instance $Z^{1}_{\;\;2}$ is the element in the first row and second column, etc. As for the fermions, we use notation $\chi^{a}_{\alpha},\; \kappa^{a,+1},\;\kappa_{b}^{-1}$. The $\kappa^{\pm 1}$ have different positions of the index, since they're in conjugate representations\footnote{These representations are equivalent, as we discussed in the text. However, we prefer to define the fermions precisely this way to get rid of some extra $\epsilon$-symbols. We should just bear in mind that the indices in this case should be contracted as $\kappa^{a,+1} \kappa_{a}^{-1}$ or $(\kappa_{a}^{\pm})^{*} \kappa_{a}^{\pm}$, etc. }. Obviously, the conjugate of a field transforms in a representation, conjugate to the one of this field. Thus, conjugation changes the position of the index. For example, $\bar{\chi}_{a}^{\alpha}\equiv(\chi^{a}_{\alpha})^{*}$, $\bar{\kappa}_{a}^{+1}\equiv (\kappa^{a,+1})^{*}$, $(Z^{*})_{a}^{\;\;b} \equiv (Z^{a}_{\;\;b})^{*}$, etc. Starting from this point, one can raise or lower indices, using $\epsilon^{ab}$ and $\epsilon^{\alpha\beta}$. For instance, $v^{a}\equiv \epsilon^{ab} v_{b}$ and $v_{a}=-\epsilon_{ab}v^{b}$. Once the minus sign in the previous formula has been written out explicitly,  $\epsilon^{ab}=\epsilon_{ab}$.

We remind the reader that $\epsilon_{ab}$ is the Clebsch-Gordan coefficient for coupling two spins $\frac{1}{2}$ to obtain spin $0$, whereas $(\epsilon \sigma_{i})_{ab}$ are the Clebsch-Gordan coefficients for coupling two spins $\frac{1}{2}$ to obtain spin $1$. This means, for instance, that $\epsilon_{ab} v_{a} w_{b}$ is a scalar, whereas $(\epsilon \sigma_{i})_{ab} v^{a} w^{b}$ is a vector.

\subsection{Fermionic charges}

The fermionic charges look the following way, when written in a manifestly covariant form:
\begin{eqnarray}\nonumber
\mathcal{Q}^{a}_{\alpha}&=&\frac{i}{4}\,\int \,d\sigma \, e^{-i\,\frac{x_{-}}{2}}\,\left(2 p_{y}\, \chi^{a}_{\alpha} +2 \epsilon^{ab}(Z^{*})^{\;c}_{b} (\epsilon_{\alpha \beta} \bar{\chi}^{\beta}_{c}+i \epsilon_{cd} \chi'^{d}_{\alpha})-\right. \\ &-& \left. \nonumber 2i \epsilon^{ab} (P_{z}^{*})_{b}^{\;c} \epsilon_{\alpha\beta} \bar{\chi}_{c}^{\beta} 
-i \epsilon_{\alpha \beta} \bar{w}^{\beta} (\kappa^{a,+1}-2i (\bar{\kappa}')^{a,-1})-i\epsilon^{ab} w_{\alpha} (\kappa_{b}^{-1}-2i (\bar{\kappa}')_{b}^{+1})+\right. \\ &+&\left. \nonumber 2 \epsilon^{ab} P_{w,\alpha}\kappa^{-1}_{b}+ 2\epsilon_{\alpha\beta} \bar{P}_{w}^{\beta} \kappa^{a,+1}-2i \,y\, (\chi^{a}_{\alpha}+i \epsilon^{ab} \epsilon_{\alpha\beta} (\bar{\chi}')_{b}^{\beta}) \right) \\
\bar{\mathcal{Q}}_{a}^{\alpha} &=& -\frac{i}{4}\,\int \,d\sigma \, e^{i\,\frac{x_{-}}{2}}\,\left(2 p_{y}\, \bar{\chi}_{a}^{\alpha} +2 \epsilon_{ab}(Z)_{\;c}^{b} (\epsilon^{\alpha \beta} \chi_{\beta}^{c}-i \epsilon^{cd} (\bar{\chi}')_{d}^{\alpha})+\right. \\ &+& \left. \nonumber 2i \epsilon_{ab} (P_{z})^{b}_{\;c} \epsilon^{\alpha\beta} \chi^{c}_{\beta} 
+i \epsilon^{\alpha \beta} w_{\beta} (\bar{\kappa}_{a}^{+1}+2i (\kappa')_{a}^{-1})+i\epsilon_{ab} \bar{w}^{\alpha} (\bar{\kappa}^{b,-1}+2i (\kappa')^{b,+1})+\right. \\ &+&\left. \nonumber 2 \epsilon_{ab} \bar{P}_{w}^{\alpha}\bar{\kappa}^{b,-1}+ 2\epsilon^{\alpha\beta} P_{w,\beta} \bar{\kappa}_{a}^{+1}+2i \,y\, (\bar{\chi}_{a}^{\alpha}-i \epsilon_{ab} \epsilon^{\alpha\beta} (\chi')^{b}_{\beta}) \right)
\end{eqnarray}
One can see that these charges are complex conjugate. They would be hermitian conjugate with respect to the Hilbert space scalar product in quantum theory.

\subsection{Bosonic charges}

Once written in covariant notation, the part of the bosonic charges quadratic in bosons looks as follows:
\begin{eqnarray}
L_{a}^{b}&=& \frac{i}{4} \int\,d\sigma\, \left( (P_{z})_{a}^{\;\;c} Z_{c}^{\;\; b}- Z_{a}^{\;\; c} (P_{z})_{c}^{\;\; b}\right)\\
R_{a}^{b}&=&\frac{i}{4} \,\int d\sigma \, \left( \bar{w}_{b} p_{w_{a}}-\bar{p}_{w_{b}} w_{a}+\frac{1}{2}\delta_{ab} \sum\limits_{i=1}^{2}\left( w_{i} \bar{p}_{w_{i}}-\bar{w}_{i} p_{w_{i}}\right)\right)\\
H&=& \frac{1}{2}\,\int \,d\sigma\,\left(\frac{1}{2} {\rm Tr}\,(P_{z}^{2}+Z'^{2}+Z^{2})+p_{y}^{2}+ y^{'\, 2} + y^{2}+\right. \\ \nonumber  &+& \left.  \sum\limits_{i=1}^{2} (p_{w_{i}} \bar{p}_{w_{i}}+w_{i}^{'} \bar{w}_{i}^{'}+\frac{1}{4} w_{i} \bar{w}_{i})\right)
\end{eqnarray}
The $U(1)$ charge is
\begin{equation}
U=\frac{i}{2}\int\,d\sigma\,\left(\bar{w}_{1} p_{w_{1}}+\bar{w}_{2} p_{w_{2}}-w_{1} \bar{p}_{w_{1}}-w_{2} \bar{p}_{w_{2}}\right)
\end{equation}
The worldsheet momentum is
\begin{eqnarray}\nonumber
p_{ws}\equiv p &=& \int\,d\sigma\, x_{-}' =-\int\,d\sigma\,\left(\frac{1}{2} {\rm Tr}\,(P_{z} Z')+\frac{1}{2} \sum\limits_{i=1}^{2}(p_{w_{i}}\bar{w}_{i}'+\bar{p}_{w_{i}} w_{i}')+p_{y} y' \right. \\ \label{pws}&-& \left.  i \bar{\chi}_{a}^{\alpha} \chi^{a\,'}_{\alpha} -i\bar{\kappa}_{a}^{+1} \kappa_{a}^{+1\,'}-i\bar{\kappa}_{a}^{-1\,} \kappa_{a}^{-1\;'})\right)
\end{eqnarray}

\subsection{Poisson brackets}

The Poisson structure can be read off, for example, from the expression for $p$ (\ref{pws}). We obtain:
\begin{eqnarray}
&\{Z_{a}^{\;\;b},\, (P_{z})_{c}^{\;\; d}\}_{P}=2 \delta_{a}^{d} \delta_{c}^{b}- \delta_{a}^{b} \delta_{c}^{d} &\\
\nonumber 
& \{w_{\alpha},\bar{p}_{w_{\beta}}\}_{P}=2 \delta_{\alpha \beta},\;\{\bar{w}_{\alpha},p_{w_{\beta}}\}_{P}=2 \delta_{\alpha\beta}, &\\ \nonumber
&\{\chi_{a}^{\alpha}, i\bar{\chi}^{b}_{\beta}\}_{P}= \delta_{a}^{b} \delta_{\beta}^{\alpha},\;\{\kappa_{a}^{+1},i\bar{\kappa}_{b}^{+1}\}_{P}=\delta_{ab},\; \{\kappa_{a}^{-1},i\bar{\kappa}_{b}^{-1}\}_{P}=\delta_{ab},&
\end{eqnarray}
all other brackets being zero.

In terms of the components of $Z\equiv z_{i} \sigma_{i}$ and $P_{z} \equiv P_{z_{i}} \sigma_{i}$ one can express the Poisson bracket of the $z_{i}$ with $p_{z_{i}}$ in the canonical form:
\begin{equation}
\{z_{i},p_{z_{j}}\}_{P}=\delta_{ij},\;\{y,p_{y}\}_{P}=1.
\end{equation}

Please note the convention of the Poisson bracket for complex fields. It is not canonical, strictly speaking, but it has been chosen in such a way that, once we write out the complex fields in terms of the real components as $w=a+i b$ and $p_{w}=p_{a}+ i p_{b}$, then $a, b, p_{a}, p_{b}$ have canonical brackets $\{a, p_{a}\}=\{b, p_{b} \} =1,\; \{a,b\}=\{p_{a}, p_{b}\}=0$. This makes it easy, for instance, to check the masses of the corresponding fields, once we plug these decompositions into the Hamiltonian.

\section{Geodesics}

As is well-known, the Penrose limit is an expansion in the vicinity of a geodesic. We call geodesics $\gamma_1$ and $\gamma_2$ equivalent, if $\gamma_2$ can be obtained from $\gamma_1$ by action of the isometry group. Since a geodesic is determined as a solution of a second order differential equation, it is determined by the initial point $\gamma(0)$ and velocity $\dot{\gamma}(0)$. Obviously, velocities $s \dot{\gamma}(0)$ define the same geodesic for any nonzero $s$ (the only difference comes from the dilation of an affine parameter on the geodesic). Thus, if $G$ acts transitively on $\mathcal{M}$ and $H$ acts transitively on $\mathrm{P}(V_\perp)$ ($\mathrm{P}$ denoting projectivization), then all geodesics are equivalent. In our case $\mathrm{P}(V_\perp)=\mathbb{R}\mathrm{P}^{5}$. A stronger condition is that, instead of the action on $\mathbb{R}P^5$, $H$ should act transitively on $S^{5}$, which might be more convenient and is probably satisfied in many cases. Another wording is that the representation of $H$ on $V$ should be irreducible over $\mathbb{R}$. For instance, this is the case for the manifold under consideration, since $V$ decomposes as $V_\perp = 3 \oplus \bar{3}$ over $\mathbb{C}$, but is irreducible over $\mathbb{R}$ under the action of $H=U(3)$. From the former viewpoint, $U(3)$ also acts transitively on $S^{5}$, which, among other things, gives rise to a coset $U(3)/U(2)=S^{5}$ (and even, cancelling the $U(1)$ factors, $SU(3)/SU(2)=S^{5}$).

There's an important exception, however, which we have omitted in the argumentation presented above. It is the case, when two geodesics 'touch' at some point $p \in \mathcal{M}$. Definition of touching is obvious and means that they both pass through the point $p$ and have the same velocity direction (once again, up to $\pm$, that is 'backward' and 'forward' are not distinguished), i.e. $\dot{\gamma}_{1}(p)\propto \dot{\gamma}_{2}(p)$. In this case, the solution of the differential equation is not specified by the point $p$ and the velocity at this point. This may well happen, since for the uniqueness of a solution a differential equation should have a regular r.h.s. (we assume that we are dealing with a system of first-order ODEs, written in the form $\dot{y}_{i}=f_{i}(\{y_{j}\})$).\footnote{For instance, consider a very simple example of equation (reminiscent of the equation of the classical giant magnon) $z'=\sqrt{z}$ with initial data $z(0)=0$. It has two solutions: $z\equiv 0$ and $z = \frac{t^{2}}{4}$. This is due to the fact that $\frac{d}{dz} (r.h.s.) = \frac{1}{2\sqrt{z}}$ is not bounded in the vicinity of $z=0$.}

For the moment we consider the question with geodesics as not totally settled, at least for us it is unclear at the moment whether any of the geodesics can touch in $\mathbb{CP}^{3}$. Of course, it should be possible to check this by a direct calculation, namely, solution of the geodesic equation.

\end{document}